\documentstyle[12pt,psfig,epsfig,aasms4]{article}
\def\degr{\hbox{$^\circ$}}
\def\arcmin{\hbox{$^\prime$}}
\def\arcsec{\hbox{$^{\prime\prime}$}}

\def\fs{\hbox{$.\!\!^{\rm s}$}}

\def\farcm{\hbox{$.\mkern-4mu^\prime$}}
\def\farcs{\hbox{$.\!\!^{\prime\prime}$}}

\begin{document}

\title{Radio and Optical Follow-up  Observations and Improved IPN
Position of  
GRB~970111}
\author{T.J. Galama\altaffilmark{1}, P.J. Groot\altaffilmark{1}, 
R.G. Strom\altaffilmark{1,2}, J. van Paradijs\altaffilmark{1,3},
K. Hurley\altaffilmark{4},
C. Kouveliotou\altaffilmark{5,6},  G.J. Fishman\altaffilmark{6},
C.A. Meegan\altaffilmark{6}, J. Heise\altaffilmark{7}, 
J.J.M. in 't Zand\altaffilmark{7}, A.G. de
Bruyn\altaffilmark{2,8}, L.O. Hanlon\altaffilmark{9},
K. Bennett\altaffilmark{10}, 
J.H. Telting\altaffilmark{11} and R.G.M. Rutten\altaffilmark{11}} 
\altaffiltext{1}{Astronomical Institute `Anton Pannekoek', University of Amsterdam,
\& Center for High Energy Astrophysics,
Kruislaan 403, 1098 SJ Amsterdam, The Netherlands}
\altaffiltext{2}{NFRA, Postbus 2, 7990 AA Dwingeloo, The Netherlands}
\altaffiltext{3}{Physics Department, University of Alabama in
Huntsville, Huntsville AL 35899, USA}
\altaffiltext{4}{Space Sciences Laboratory, Berkeley CA 94720-7450, USA}
\altaffiltext{5}{Universities Space Research Asociation}
\altaffiltext{6}{NASA/MSFC, Code ES-84, Huntsville AL 35812, USA}
\altaffiltext{7}{Space Research Organization Netherlands, Sorbonnelaan 2,
3584 CA Utrecht, The Netherlands}
\altaffiltext{8}{Kapteyn Astronomical Institute, Postbus 800, 9700 AV,
Groningen, The Netherlands} 
\altaffiltext{9}{Physics Department, University College Dublin,
Belfield, Dublin 4, Ireland} 
\altaffiltext{10}{Astrophysics Division, Space Science Department of
ESA, ESTEC, 
Noordwijk, The Netherlands}  
\altaffiltext{11}{ING Telescopes/NFRA, Apartado 321, Sta. Cruz de La
Palma, Tenerife 38780, Spain} 

\begin{abstract}
We report on Westerbork 840 MHz, 1.4 and 5 GHz radio observations of
the improved IPN--WFC error box of the 
$\gamma$-ray burst GRB~970111, between 26.4 hours and 120 days
after the event onset. In the $\sim$ 13 sq
arcmin area defined by the IPN (BATSE and {\it Ulysses}) annulus and the 
published refined BeppoSAX Wide Field Camera (WFC) error box we 
detected no steady sources brighter than 0.56 mJy ($4\sigma$), and no 
varying radio emission, down to 1.0 mJy (4$\sigma$). 
We also report on $B, V, R$ and $I$ band
observations of the error box with
the 4.2 m William Herschel Telescope at La Palma. 
\end{abstract} 

\keywords{gamma rays: bursts --- gamma rays: individual (GRB
970111) --- radio continuum: general}


\section{Introduction}

The `isotropic' and `inhomogeneous' distribution of gamma-ray bursts (GRBs) 
(Meegan et al. 1992; Briggs et al. 1996) is explained in a natural way
if GRB sources are 
located at cosmological distances of order Gpc (Usov and Chibisov 1975; 
Paczy\'{n}ski 1986; Goodman 1986; 
Fenimore et al. 1993).
However, a GRB source distribution in a very large
galactic halo (characteristic size $10^5$ pc) is also advocated
(see, e.g., Podsiadlowki et al. 1995). 
Independent of what the distance scale of GRB sources is,   
the mechanism that gives
rise to the sudden emission of gamma rays is unknown, and it is
generally believed that progress in solving this problem requires
the detection of GRB counterparts in other parts of the
electromagnetic spectrum. Very promising in providing new
insights to the origin and production mechanism of GRBs are the recent
discovery of an optical transient 
related to GRB 970228 (Van Paradijs et al. 1997; Groot et al. 1997a),
and the monitoring of its  optical light curve (Galama
et al. 1997a; Sahu et al. 1997). 
The optical transient may be associated with a faint galaxy
(Van Paradijs et al. 1997; Groot et al. 1997b), 
suggesting that it lies at a cosmological distance. 

The peak luminosities of GRBs are highly super-Eddington, even for
galactic halo models.  
Paczy\'{n}ski and Rhoads (1993) pointed out that it is hard to avoid
relativistic outflows from the GRB source, and that the interaction of
these flows with an interstellar medium must generate radio emission,
as is observed, e.g., in extragalactic jet sources (see also Katz
1994b; M\'{e}sz\'{a}ros et al. 1994,  M\'{e}sz\'{a}ros and Rees
1997). Paczy\'{n}ski and Rhoads estimated 
that the strongest GRBs may be followed by transient ($\sim 20$ mJy)
radio emission at intervals ranging between minutes (for
the galactic-halo distance scale) to roughly weeks to months (for the
cosmological distance scale).

We report here on observations of the refined error box of GRB~970111
(In 't Zand et
al. 1997) made with the Westerbork Synthesis Radio Telescope (WSRT) in
the months following the burst, and the derivation of a substantially improved
Interplanetary Network (IPN) error box based on the time delay between
the burst as observed 
with the Burst And Transient Source Experiment (BATSE) and {\it Ulysses}. In
Section 2 we describe the properties of the 
GRB as observed with BATSE, and present the IPN-BeppoSAX error box. In 
Section 3
we describe the radio and optical observations, the results of which 
are discussed in Section 4.

\section{GRB~970111 gamma-ray observations}

\subsection{BATSE observations}

\begin{figure}[p]
\centerline{\psfig{figure=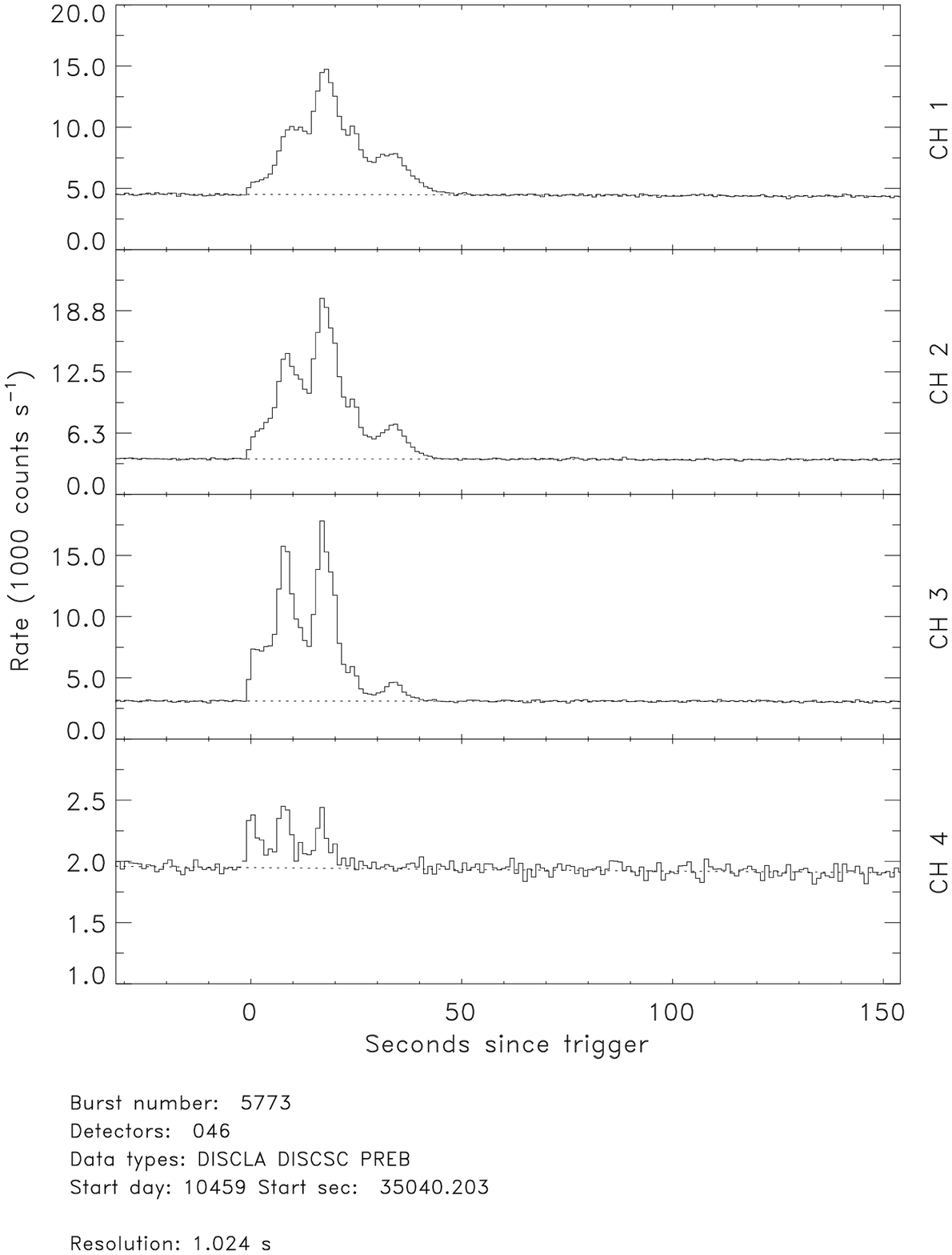,bbllx=100,bblly=150,bburx=580,bbury=700,clip,width=8cm}}
\figcaption[]{BATSE light curve of GRB~970111 in the four LAD energy
channels, 25-50, 
50-100, 100-300 
and  $>$300 keV (channel 1 being that of the lowest energy). Time resolution is
1.024 s. \label{BATSE}}
\end{figure}

GRB~970111 triggered
BATSE on 11 January, 1997 at 09:44:00 UT. Here we describe data obtained 
with the Large Area Detectors (LADs); a detailed description of
the instrument is given elsewhere (Fishman et
al. 1989). Fig. \ref{BATSE} shows the  
burst profile (with 1.024 s time resolution) in four energy channels 
(25-50, 50-100, 100-300, $>$300 keV). 
The event duration is $T_{90}=31.7$ s, estimated
from the summed signal of all energy 
channels (T$_{90}$ is the time during which 90\% of the total burst
counts are 
emitted; Kouveliotou et al. 1993). 

When we fit a broken power law to the event energy spectrum we find that the
total 25 keV - 1.8 MeV fluence during 48.4 s of emission is 
$(5.8\pm 0.3)\times 10^{-5}$
ergs cm$^{-2}$. This fluence is among the top 1\% of all GRBs observed
with BATSE so 
far. The event peak flux (summed over the same energy range in 1.024 s
interval) is $(4.4 \pm 0.2)\times 10^{-6}$ ergs cm$^{-2}$s$^{-1}$, which
makes GRB970111 one of the top 5\% 
brightest BATSE events. A slightly better fit is obtained using the Band
function (Band et al. 1993), for which we find that the
spectral power peaks at $E_{\rm peak}$ = 166$\pm$1 keV. 

In the burst profile summed over all channels we distinguish five peaks: 
channel 4 shows only the first three, channel 1 shows only the 
last four of them. The peaks occur at times (since trigger) $T \sim 1$~s, 
$\sim 8$~s, $\sim 17$~s, $\sim 25$~s, and $\sim 35$~s. 
Thus, this event shows strong spectral evolution, with the first peak 
having a very hard spectrum starting above $\sim 100$~keV, and the last two 
having much softer spectra. 
This is confirmed in the BeppoSAX Wide Field 
Camera observations (WFC; energy range 1.8-30 keV) in which there are only two 
peaks starting $\sim 20$ and $\sim 35$~s,
respectively, after the BATSE trigger time.
Spectral evolution in GRBs is well known; GRB~970111 is an extreme
example of late pulses disappearing entirely at high energies
(see Ford et al. 1995 for a 
detailed description of spectral-evolution trends of bursts observed
with BATSE).  

The burst location using the directional response of the BATSE detectors that
recorded the burst was 
$\alpha=231^{\circ}$
and $\delta$$=19^{\circ}$ (J2000) with a total error of
1.5$^{\circ}$ (statistical and 
systematic added in quadrature). Fortunately, besides the two near-Earth 
satellites CGRO and BeppoSaX, the event 
was also recorded with the {\it Ulysses} spacecraft, which allowed 
the derivation of a very narrow and short IPN annulus (see Hurley et
al. 1994 for a description of IPN error boxes).

\subsection{IPN--BeppoSax WFC combined error box}

Triangulation of this event using {\it Ulysses} and BATSE data gives an
annulus of possible arrival directions.  This annulus is centered
at $\alpha =11^{\rm h}$50$^{\rm m}$55.4$^{\rm s}$, 
$\delta$ +33\degr21\arcmin57\arcsec (J2000), and 
has a radius of $49.983 \pm 0.022^{\circ}$ (3$\sigma$ confidence).   
It is consistent
with the one given by Hurley et al. (1997a),
but narrower.  
Combining this annulus with the refined BeppoSAX WFC error
circle (In 't Zand et al. 1997) 
gives an approximately trapezoidal error box whose corners are at
$\alpha$, $\delta =15^{\rm h}$28$^{\rm m}$22$^{\rm s}$, 
$+19\degr33\arcmin45\arcsec ; 15^{\rm h}$28$^{\rm m}$28$^{\rm s}$, 
$+19\degr36\arcmin28\arcsec ; 15^{\rm h}$28$^{\rm m}$09$^{\rm s}$, 
$+19\degr33\arcmin40\arcsec ; 15^{\rm h}$28$^{\rm m}$21$^{\rm s}$, 
$+19\degr38\arcmin59\arcsec$
(J2000), and whose area is approximately 13
square arcminutes (see Fig. \ref{GRBfield}).  Of the three SAX and
ROSAT quiescent soft 
X-ray sources (Butler et al. 1997; Voges
et al., ibid) found in the preliminary WFC error circle (Costa et
al. 1997) none lies within this error box.  

\section{Radio and Optical Observations}
\subsection{Radio Observations}

The error box of GRB~970111 was
first observed at 840 MHz with the WSRT at the preliminary position given by
Costa et al. (1997) on 
January 12, at UT 12:51, starting 26.4 hours after the event, for only 1.44
hours until the GRB source exceeded the telescope hour-angle
limit. Subsequent 
12 hour observations were made on January 13.31, 14.31, 16.31, 20.30
and 29.27, March 
9.16, and April 9.08 UT (these and subsequent observing times refer to
the middle of the 
observing period).


We used a new UHF receiving system, whose 
upper frequency band covers the 700--1200 MHz range, and 
selected 840 MHz because electromagnetic interference
(EMI) is usually absent there. The standard continuum correlator was used,
providing us with eight contiguous 10 MHz bands from 800 to 880 MHz. The
noise level in a continuum map for a 10 MHz bandwidth after 12 hours of
integration is typically 0.5 mJy beam$^{-1}$, and our analysis shows that
this was usually achieved in the present series of measurements. At 840 MHz,
the synthesized beamwidth at $20^\circ$ declination is
$21^{''}\times62^{''}(\rm
RA \times Dec)$, and the field of view (half-power beamwidth) is about
$1^{\circ}$. 

We also observed at 5 GHz on each of February 26.05 (5.15 hrs) and
27.29 (5.55 hrs) and March 2.22 (10.24 hrs).
With 8
bands of each 10 MHz width we obtained noise levels $\sim$ 0.15 mJy per
observation. The synthesized beamwidth at $20^\circ$ 
declination is
$6^{''}\times18^{''}(\rm
RA \times Dec)$. The field of view (half-power beamwidth) is about
$0.17^{\circ}$.

The latest observations were each for 8.4 hours at 1.4 GHz on May 11.07
and 12.06 (120.6 
days after the event). The 1.4 GHz data consist of 5 bands of width 10
MHz (two were bad) and 3 bands of width 5 MHz, providing a useful 45 MHz of
bandwidth. The noise 
level combining the two 1.4 GHz observations is 0.070 mJy. The
synthesized beamwidth at $20^\circ$  
declination at 1.4 GHz is $13^{''}\times37^{''}(\rm
RA \times Dec)$. The field of view (half-power beamwidth) is about
$0.6^{\circ}$.

The data were analyzed using the NEWSTAR software 
package\footnote{http://www.nfra.nl:80/newstar/}.  
The interferometer complex visibilities were first examined for
possible EMI and other obvious defects. Bands with strong interference
were either deleted or carefully edited. Due to interference about 45
\% of the data were 
unusable in the January 14 observation. On the other days 
interference was usually limited to about 10 \%.
In the self calibration (gains and phases) we used models for each of the 
8 bands around 840 MHz, each containing nearly 80 discrete sources above 2.5
mJy. For calibration of interferometer gain and
phase, for each 
observation and each band, we used the standard WSRT calibrator 3C 48
(24.05 Jy at 840 MHz, 15.96 Jy at 1.4 GHz and 5.55 Jy at 5 GHz; based
on the 3C286 flux density on the Baars et  
al.\ 1977 scale). 
The 840 MHz map of the field produced by combining data from January 13, 14, 16
20, 29 and March 9 is shown in Fig. \ref{GRBfield}.
The 1.4 GHz observations have been analyzed in a similar way, but no self
calibration was performed. 
The 5 GHz observations contained only very weak
sources within the field of view and hence no model has been obtained
and no self calibration was performed.

We searched for a transient radio counterpart by looking for
variability at 840 MHz within
the refined WFC-{\it Ulysses}-BATSE error region. 
Since all our observations were obtained with the same
configurations of the Westerbork array, the synthesized beams of the
images for different days are in principle the same. 
The most sensitive way to look for variations
in the data is via the construction of difference
maps from residual maps, i.e. maps deconvolved with a model
containing all real sources in the field. 
We obtained equal synthesized beams in
the residual maps of two observations by retaining only those
$u-v$ points in common to the datasets of any two days in the comparison. 
Subsequently, we
subtracted the two maps from each other and obtained the difference
map. In these
difference maps we did not detect any source variation in the
13 square arcminute error box of GRB~970111, down to 1.0
mJy (4$\sigma$) for the period of 26.4 hours to 88 days after the burst,
and down to 2.4 mJy
(4$\sigma$) for the first observation (January 12; see
Fig. \ref{Strom}).

Since no radio source is found in the error box at any of the
observing  frequencies  (840 MHz,
1.4 and 5 GHz) the 1.4 and 5 GHz upper
limits to variability have been determined by considering the limit
for detection of 
a radio source (for which we take 4$\sigma$, corresponding to 0.28 mJy
at 1.4 GHz and 
0.6 mJy at 5 GHz; see
Fig. \ref{Strom}).

\subsection{Optical Observations}

We made $B$, $V$, $R$ and $I$ band images of the refined
error box of GRB 
970111 
using the 4.2m William
Herschel Telescope (prime focus) on La Palma in service time. 
On Feb. 28.2 UT $B$, $V$ and $I$ images were obtained with integration
times of 930, 405, and 
695 s, respectively, and an $R$ image (405 s) on Mar. 1.1 UT.
The field of view is 
7\farcm2 $\times$ 7\farcm2. Fig.\ \ref{fig:wht} shows a $V$-band image of the region
covering the WFC error box of GRB 970111. We made a $B, V$ and
$I$ magnitude
calibration with images of the photometric 
Landolt standard field 107 using stars 599-602. For the $R$ magnitude
calibration we used Landolt standard field
104, star 334, 336, 338 and 339. Comparison with the Digital
Sky Survey shows no brightening of 
sources to V = 19. The two radio sources located near to the error box
(Sect. \ref{Disc}) show no counterpart down to $B$ = 23.7, $V$ = 24.0,
$R$ = 22.9 and $I$ = 21.7. The 
combined GRB error box contains a few galaxies above the detection
limit. One is a relatively bright galaxy (see Fig. 3), 
at $\alpha$ = 15$^{\rm h}28^{\rm
m}$15\farcs2, $\delta$=
+19$^{\circ}35^{'}55^{"}$ (equinox 2000.0), with $V$ = 20.2 $\pm$ 0.1,
$B-V$ = +1.3 $\pm$ 0.1, 
$V-R$ = +0.7 $\pm$ 0.1, 
$V-I$ = +1.4 $\pm$ 0.1. This galaxy is not 
detected in the radio observations at 840 MHz ($<$ 0.42 mJy,
3$\sigma$),
1.4 GHz ($<$ 0.21 mJy, 3$\sigma$) and 5 GHz
($<$ 0.45 mJy, 3$\sigma$).

\section{Discussion \label{Disc}}

\begin{figure}[p]
\centerline{\psfig{figure=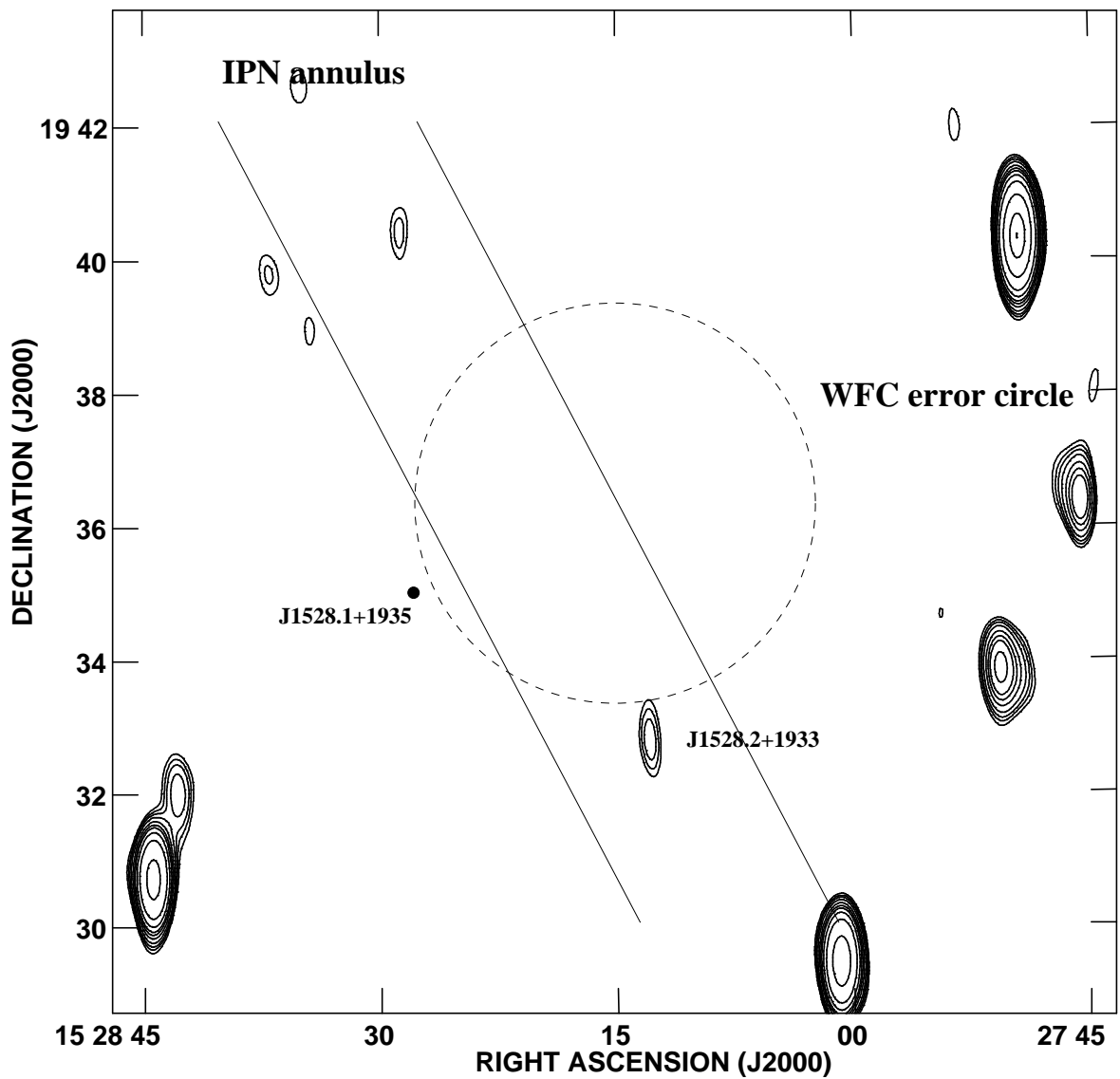,width=16cm}}
\figcaption[]{Contour plot of the WSRT 840 MHz image  
of the GRB 970111
field  centered at the refined WFC position 15 $^{\rm h}$ 28$^{\rm m}$
15$^{\rm s}$, 
+19 $^{\circ}$ 36 $^{'}$
18 $^{''}$ (equinox J2000). Combined are data from January 13, 14, 16
20, 29 and March 9. Included are  
the refined WFC 3$^{'}$ radius error region ($3\sigma$) and the
refined $3\sigma$ 
{\it Ulysses}-BATSE
triangulation annulus. Indicated is the radio source J1528.2+1933, and
the position (filled circle) of the radio source J1528.1+1933, both
located just 
outside the WFC error box.  
Contour levels are 0.56, 0.70, 0.84, 1.05, 1.40, 1.75, 2.10, 2.80,
4.20, 8.40, 16.80 and 22.25 mJy (0.14 mJy noise). 
\label{GRBfield}} 
\end{figure}

Using the combined map based on observations made on 
12, 13, 16, 20, 29 January and March 9, we find no 840 MHz radio 
sources in the error box of GRB~970111 (see Fig. \ref{GRBfield}), down 
to the detection limit of 0.56 mJy (4$\sigma$). 
The ($4\sigma$) detection limits at 1.4 GHz and 5 GHz are 0.28 mJy 
and 0.6 mJy, respectively. (Note that the limits reported in Section 3.1 
refer to variability of any radio source in the error box). Two sources lie
just outside the error box at $\alpha$, $\delta$ = 15$^{\rm h}28^{\rm
m}$12\fs8 $\pm$ 0\fs1, +19$^{\circ}32^{'}50^{''}$ $\pm 4^{''}$
(J1528.2+1933) and $\alpha$, $\delta$ = 15$^{\rm 
h}28^{\rm m}$04\fs9 $\pm$ 0\fs1,
+19$^{\circ}35^{'}00^{''}$ $\pm$ 2$^{''}$ (J1528.1+1935). The 840 MHz, 1.4 
and 5 GHz 
fluxes are as follows: 1.13 $\pm$ 0.14 mJy, 0.275 $\pm$ 0.07 mJy 
and $<$ 0.45 mJy (J1528.2+1933), and 
$<$ 0.42 mJy, $<$ 0.21 mJy and 0.80 $\pm$ 0.15 (J1528.1+1935)
(upper limits 
are 3 $\sigma$). The source A reported by
Galama et al. (1997b) appeared, after the final analysis of 
subsequent observations, to be a noise peak. 

Previous follow-up radio observations have not led 
to the detection of a variable radio signal that is 
possibly correlated  with a GRB (see, e.g., the recent review by 
Vrba 1996). The deepest radio follow up at short
intervals so far is that by 
Frail et al. (1994) who obtained $5\sigma$ upper limits of 3.5 mJy 
to variability at 1.4 GHz , within 3 to 99 days of 
GRB 940301, from observations of the COMPTEL error box
of this burst (Frail et al. 1994). Galama et al. (1997c) derived upper
limits of 10 mJy to variability at 325 MHz during an interval of 1 to
3 months after GRB 940301. 
Our present results improve the response time of Frail et al. (1994)
by a factor $\sim$ 3. A similar improvement 
has been reported by Frail et al. (1997a).

To interpret our null result, we take as a guideline 
the work of Paczy\'{n}ski and Rhoads,  which is based on the idea that GRBs 
are caused by the interaction of a relativistically expanding fireball with 
the interstellar medium 
(with density $\rho = 10^{-24}\rho_{24}$ g~cm$^{-3}$).
From their results (see their Eqs. 18b and 19b) 
we estimate, using a burst fluence $S = 5.8 \times 10^{-5}$ erg~cm$^{-2}$
and an observing frequency of 840 MHz, that the peak radio flux
$F_{\rm max}$ and
the time $t_{\rm max}$ since the GRB that this peak is reached are given by: 
$F_{\rm max} \sim (4.1\, {\rm mJy})~d^{-1/4}_{0.5}~\rho^{1/8}_{24}$ 
and 
$t_{\rm max} \sim (75\, {\rm days})~d_{0.5}~\rho^{1/2}_{24}$; here $d_{0.
5}$ is the source distance in units of 0.5 Gpc.
(In these estimates we have set several of the dimensionless
constants occurring in these equations equal to unity). 
During the rise of the radio transient the flux increases with time
$t$ proportional to $t^{5/4}$; the decay goes as $t^{-3/4}$.

The model of Katz (1994a,b) predicts an initial 840 MHz flux 
(simultaneous with the GRB, taken to emit $5.8 \times 10^{-5}$ ergs 
cm$^{-2}$ in $\sim$ 30 s, over a 4 $\times 10^{19}$ Hz bandwidth
centered at $4 \times 
10^{19}$ Hz) of 
$\sim 1.3$ $\mu$Jy.  This flux rises proportionally to $t^{4/5}$ until 
it reaches a
peak, roughly 30 s $\times$ (840 MHz/4 $\times$ 10$^{19}$
Hz)$^{-5/12}$ $\sim$ 10 days 
later, around 5 mJy.  Thus, the models of Paczy\'nski
and Rhoads and of Katz predict radio afterglows for GRB 970111 that 
are quite similar.

GRB~970111 was extremely strong: it was among the top 1\% of BATSE bursts 
with respect to fluence, and the top 5\% with respect to peak flux. 
In the cosmological model for GRBs the turnover in 
the cumulative burst peak flux distribution occurs at redshift 
$z \sim 1$ (e.g., 
Fenimore et al. 1993). Since the luminosity distribution of 
observed bursts is not very broad (see, e.g., Horack et al 1994; 
Ulmer et al. 1995) 
this indicates that GRB~970111 originated at a rather modest redshift (of 
order 0.1). We will therefore, in the following, assume 
a nominal source distance of 0.5 Gpc.

\begin{figure}
\centerline{\psfig{figure=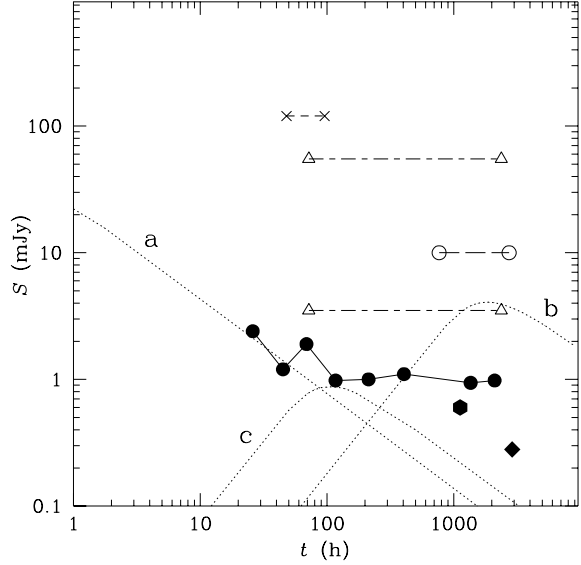,width=8cm}}
\figcaption[]{Log -- log plot of the upper limits to variable radio
emission from GRB 970111 
(filled symbols) and GRB 940301 (crosses, open circles and open triangles) 
as a function
of elapsed time since the gamma-ray burst. The flux density limits at 840 MHz
(filled circles) are from the eight WSRT observations discussed in this 
paper (GRB 970111). The other limits are for 1.4 GHz (solid diamond) and 5 
GHz (solid hexagon). The remaining measurements all
refer to GRB 940301, at 151 MHz 
(Koranyi et al., 1995; crosses), 327 MHz (Galama et al., 1997; open circles), 
408 MHz (open triangle, upper) and 1400 MHz (open triangle, 
lower; Frail et al., 1994),
all being upper limits to candidate sources in two or more observations over
the time span indicated by the horizontal lines. The three dotted curves show
the expected radio emission at 840 MHz from a gamma-ray burst with a fluence
of $5.8\times10^{-5}\rm\;ergs\,cm^{-2}$ according to the calculations of
Paczy\'nski and Rhoads (1993). Curve {\bf a} is for a distance of 100 kpc
(and peaks at 34 mJy after $t=22$ minutes), {\bf b} for 0.5 Gpc, both for
nominal values of ambient density, $\rho=10^{-24}\rm\;g\,cm^{-3}$, and
efficiency, $\xi=0.1$ (see text). Curve {\bf c} is for 0.5 Gpc, with 
$\rho=1.0\times10^{-26}\rm\;g\,cm^{-3}$ and $\xi=0.3$ \label{Strom}} 
\end{figure}

\begin{figure}
\centerline{\epsfig{figure=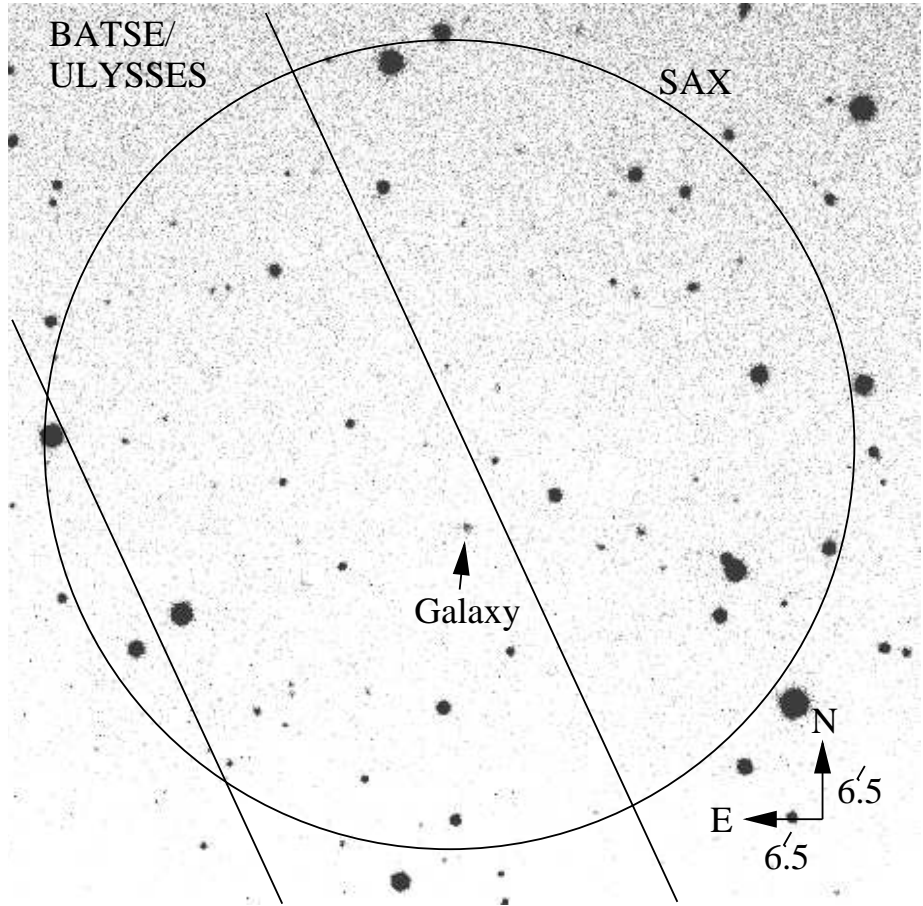,width=12cm,angle=-90}}
\figcaption[]{WHT prime focus V band image of the field of GRB 970111. 
Indicated are the Beppo SAX WFC error circle, the IPN annulus, and the 
relatively bright galaxy mentioned in the text. \label{fig:wht}}
\end{figure}

In Fig. \ref{Strom} we have plotted our upper limits as a function of
time since the 
burst, together with some of the best previous limits, and three curves showing
the radio behaviour according to the above expressions derived from the
Paczy\'nski and Rhoads model.
It appears that our observations
exclude the model of Paczy\'nski and Rhoads at its nominal parameters (the 
same is the case for the Katz model). They suggest that for the 
model to be viable, GRB 970111 should either have come from 
very nearby ($<20$ kpc, peaking within
a few minutes of the burst), or from a very large distance 
($>3$ Gpc, with a peak after many
months, in which case its peak luminosity would have been exceptionally 
high), or it occurred with a high efficiency of gamma-ray production in
regions of very low ambient density. 

Although our non-detection of a radio counterpart for GRB 
970111 nominally excludes the models of Paczy\'nski and Rhoads  (1993) and of 
Katz (1994a,b), we feel that in both models there are enough 
scaling factors that disagreement by about one order of magnitude is not 
yet compelling evidence that the models are qualitatively wrong. 
Moreover, at low radio frequencies the emission may be limited by
self-absorption (e.g. Katz 1994a).

After the revised version of this paper was near completion we learnt 
of the detection of an optical and  radio counterpart of GRB970508
(Bond 1997; Frail et al.  
1997b). The inverted spectrum $F_{\nu} \propto \nu^{+1.1}$ observed for
this radio counterpart is evidence for the occurence of self-absorption.
The measurement of a redshift ($z$ $>$ 0.835, Metzger et
al. 1997) of the optical counterpart of GRB 970508 may have
settled the distance scale to GRB sources.
It is clear that the relative response in radio is far 
from proportional to the strength of the GRB; the same is the case for 
the optical response 
(Castro-Tirado et al. 1997). It is currently unclear what GRB 
characteristic determines 
the low-energy response to GRBs, and more follow-up observations are
required to unravel their systematics.

We will continue monitoring the error box
of GRB 970111 in an attempt to detect any delayed radio signal from this 
event.

\acknowledgments

We are grateful for the assistance of the WSRT telescope operators
G. Kuper and R. de Haan. The WSRT is operated by 
the Netherlands Foundation for Research in Astronomy (NFRA) with
financial aid by the Netherlands Organization for Scientific Research
(NWO). The WHT is operated on behalf of the English PPARC and Dutch
NFRA at the Spanish observatory Roque de Los Muchachos on La Palma,
Spain. 
T. Galama is supported through a grant by NFRA under contract
781.76.011. C. Kouveliotou acknowledges support from NASA grant NAG 5-2560.
K. Hurley acknowledges support for {\it Ulysses} under JPL contract 958056,
and for the IPN under NASA grant NAG 5-1560.

\newpage

\newpage

\end{document}